\documentclass[aps,prd,preprintnumbers]{revtex4}
\usepackage{graphics}
\def \beq{\begin{equation}}
\def \eeq{\end{equation}}
\def \beqa{\begin{eqnarray}}
\def \eeqa{\end{eqnarray}}

\def\ie{{\sl i.e.\/}}
\def\x{\mathbf x}

\def\annp{{\sl Ann.\ Phys.\/}}

\def\npa{{\sl Nucl.\ Phys.\/}, A}

\begin{document}
\title{A systematic expansion for relativistic causal hydrodynamics}
\author{Sourendu \surname{Gupta}}
\email{sgupta@tifr.res.in}
\affiliation{Department of Theoretical Physics, Tata Institute of Fundamental
         Research,\\ Homi Bhabha Road, Mumbai 400005, India.}

\preprint{TIFR/TH/07-20}

\begin{abstract}
A systematic expansion of the Boltzmann equation for the diffusion
of dilute tracers, in powers of the Knudsen number, carried out to
next-to-leading order (NLO), gives the relativistic causal diffusion
equation (Kelly's equation).  Using dimensionless combinations of
dynamical quantities, we show when the small NLO term plays a crucially
important role.  We proceed to show that a derivation of Kelly's equation
from a microscopic theory of the correlation function of the number
density of diffusers is possible. The correlator fulfils the Green-Kubo
relation for the diffusion constant, as well as an f-sum which goes beyond
a purely phenomenological causal theory. We argue that the construction
generalizes to the full hydrodynamic equations.
\end{abstract}
\maketitle

After Eckart \cite{eckart} and Landau \cite{landau} cast Navier-Stokes
hydrodynamics into manifestly Lorentz covariant form, it was realized
that there still remained an essential acausality in the theory
\cite{acausal,kelly}.  The problem was solved in principle by Israel and
Stewart \cite{causal} who wrote down a phenomenological theory which has
since been applied to heavy-ion collisions, the quintessential laboratory
for ultra-relativistic hydrodynamics \cite{hic}.  The hydrodynamic
equations are obtained by making a derivative expansion of the transport
equations. If the longest microscopic time scale in the problem is a
relaxation time $\tau_R$, then the hydrodynamic equations are valid
for times, $\tau\gg\tau_R$. When the Knudsen number, $K=\tau_R/\tau$,
goes to zero, the microscopic origin of the theory is effaced and what
remains is a continuum theory\footnote{In non-relativistic gas dynamics,
the Knudsen number, also called the continuum parameter, is defined to
be the ratio of a mean free path length to a typical length scale in
the problem. In the relativistic context a ratio of time scales is more
appropriate.}. The usual equations of hydrodynamics are obtained in this
way at leading order (LO) in $K$, \ie, at ${\cal O}(K^0)$ \cite{grad}.
Here we discuss the version of the equations which appears at the
next-to-leading order (NLO) in $K$.

Two questions seem not to have been addressed fully in the literature,
and are interesting enough that we consider them briefly here. The
first question is the following: is the restoration of causality at
NLO necessary, or is it an accident?  The second question is: given a
correlation function computed from the underlying microscopic theory,
how does one show that the causal relativistic hydrodynamic theory
is obtained at late times?  We examine these questions here for the
simplest of hydrodynamic equations--- the diffusion equation, and its
causal version: Kelly's equation. The answers are generalizable to all
hydrodynamic equations, as we argue.

\begin{figure}[htb]\begin{center}
   \scalebox{0.707}{\includegraphics{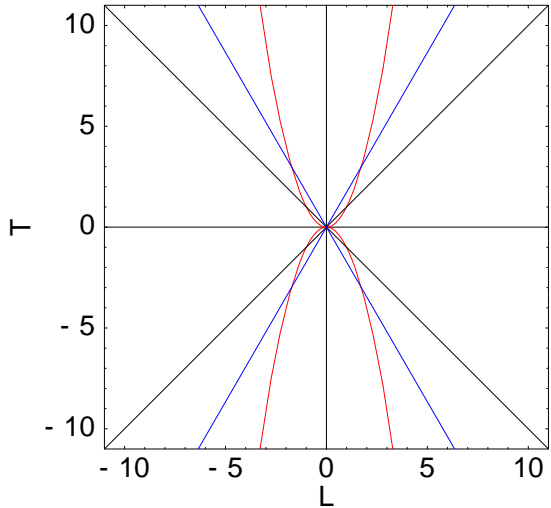}}
   \end{center}
  \caption{The parabola is a locus of constant $\cal W$ and contains the
    diffusive wake. The sets of straight lines are loci of constant $M^2$ (the
    inner set for $M=1$, and the outer for $M=1/c_s$, \ie, the light
    cone) and denote the causal front. The lack of relativistic causality of
    the diffusion equation is seen in the fact that the parabola lies
    outside the light cone at small $\tau$. For any $M$, the front and wake
    can be distinguished at late times.}
\label{fg.causality}\end{figure}

Recall that the continuity equation for the density, $n$, of a dilute tracer
inserted into a thermalized background fluid,
\beq
   \frac{\partial n(\x,t)}{\partial t} + \nabla\cdot\mathbf J(\x,t) =0,
\label{continuity}\eeq
where $\mathbf J$ is the tracer current, expresses conservation
of the number of tracer particles. Putting this together with the
phenomenologically established Fick's law,
\beq
   \mathbf J(\x,t) = - D \nabla n(\x,t),
\label{fick}\eeq
where $D$ is called the diffusion constant, gives rise to
the diffusion equation,
\beq
   \frac{\partial n(\x,t)}{\partial t} = D \nabla^2 n(\x,t).
\label{diffeq}\eeq
$D$ is an example of a transport coefficient.
A comparison of the strength of the two terms in the equation yields
a dimensionless variable which we call the Wiener number,
\beq
   {\cal W} = \left(\frac{\partial n}{\partial t}\right)
      \biggm/ \left(D \nabla^2 n\right) = \frac{L^2}{\tau D},
\label{comps}\eeq
where $\tau$ is the time and $L$ the distance over which
the diffusion phenomenon is observed. For typical diffusive behaviour
we expect ${\cal W}$, to be finite, non-zero, and constant in time,
which yields the familiar result $L\simeq\sqrt{D\tau}$. This violates
relativistic causality for $\tau<D{\cal W}/c^2$, since $L>c\tau$ in
that case.  This is shown in Figure \ref{fg.causality}.  One suspects
that the diffusion equation is not valid at very early times, but there
is nothing in the equation that prevents this.  As is well-known, the
local conservation equation (eq.\ \ref{continuity}) obeys relativistically
causality. As a result, its violation in the eq.\ (\ref{diffeq}) can be
traced to Fick's law.

Kelly \cite{kelly} examined the Boltzmann transport equation for
the distribution function of a dilute gas of tracer particles
in the background of a thermal gas in equilibrium. The transport
theory is assumed to satisfy relativistic causality. The condition
that the background gas has no flow selects out a special frame in
which the gas is at rest. In the remainder of this note we work in
this frame, giving up manifest Lorentz covariance for notational
ease\footnote{Manifest covariance can be reinstated by using the
time-like velocity 4-vector $u_\mu$ and replacing time derivatives by
$D=u^\lambda\partial_\lambda$. Spatial derivatives are then the orthogonal
projections.}. The diluteness aproximation is used to retain only linear
terms in the tracer density inside the collision integral. The collision
term then yields up a natural microscopic scale of time through the
collision frequency, $\nu=1/\tau_R$.

It was shown that the expansion carried out to order $K$ yielded the
hyperbolic diffusion equation---
\beq
   \tau_R\frac{\partial^2 n}{\partial t^2}
     + \frac{\partial n}{\partial t} = D \nabla^2 n,
\label{reldiffeq}\eeq
and, an expression for the diffusion constant, 
\beq
   D = \tau_R c_s^2,
\label{dexpr}\eeq
where $c_s^2$ is the RMS velocity of the tracer particles.  Explicit
solutions of the differential equation with this subsidiary condition
show that relativistic causality is restored \cite{kelly}.

The relative importance of the causal (the second derivative in energy)
and the frictional terms (the first derivative) is given by the dimensionless
ratio
\beq
   \left(\tau_R\frac{\partial^2 n}{\partial t^2}\right)\biggm/
   \left(\frac{\partial n}{\partial t}\right) \simeq \frac{\tau_R}\tau = K.
\label{kfind}\eeq
This is exactly as expected, since the causal term arises at
NLO in $K$, and one obtains the diffusion equation by neglecting it.
The relative importance of the causal term and the force is
\beq
   M^2=\left(\frac{\tau_R}D\right)\,\left(
      \frac{\partial^2 n}{\partial t^2}\right)\biggm/\bigl(\nabla^2n\bigr)
    \simeq \left(\frac{L}{c_s\tau}\right)^2.
\label{rdef}\eeq
Phenomena occurring at constant $M^2$ would include a causal diffusive
front propagating with speed $V=L/\tau$.  Thus $M$ is the diffusive
analogue of the Mach number.  The three dimensionless constants are not
independent, since, by construction, $M^2=K{\cal W}$.

The existence of a non-vanishing scale $\tau_R$ allows restoration of
relativistic causality. On the light cone one has $L=c\tau$. The locus of
constant $\cal W$ lies completely inside the light cone for a $\tau_*$
given by $\sqrt{{\cal W}D\tau_*}\le c\tau_*$.  A little manipulation
using the expression for $D$ in eq.\ (\ref{dexpr}) brings this to the
form $K_*=\tau_R/\tau_*\le M_c^2/{\cal W}$, where $M_c$ is the value
of $M$ when $V=c$ and $K_*$ is the value of $K$ at which causality is
restored. One may note, with \cite{kelly}, that for $\tau<\tau_*$ the
frictional term in eq.\ (\ref{reldiffeq}) can be neglected, so that using
eq.\ (\ref{dexpr}), the remaining terms become a wave-equation. Since
$c_s^2<c^2$, relativistic causality is restored.  More correctly, one
notes that in the problematic region $K>K_*\simeq{\cal O}(1)$, and hence
the expansion breaks down. Thus, causality is restored by shielding
the problematic region from the continuum equation.  Investigation of
$\tau<\tau_*$ must be carried out in transport theory, which is causal
by construction.

In \cite{kelly} it was noted that the solutions of eq.\ (\ref{reldiffeq})
consist of two distinct regions--- the ``wake'', \ie, $L\propto\sqrt\tau$,
where characteristic diffusive behaviour is seen, and the ``front'', \ie,
$L\propto\tau$, where causal behaviour is restored. This is precisely
the result of our analysis here (see Figure \ref{fg.causality}). The
wake is characterised by the physics of fixed $\cal W$, and the front by
fixed $M$. $\cal W$ is, of course, inherent in the continuum theory. The
appearance of a new, non-vanishing, dimensionless constant $K$ implies
the existence of a new constant $K{\cal W}$, which leads to a shielding of
the acausal region, as discussed above. This answers our first question:
the restoration of causality is a necessary consequence of the NLO
expansion in $K$.

This analysis also tells us when to use Kelly's equation rather than the
usual diffusion equation. Clearly, at late times, the differences between
the two are minor in the region of the wake.  The strength of the front
decays exponentially in $1/K$ \cite{kelly}, so this is also negligible
at late times. At late times, therefore, the physics of the two equations
is identical. Kelly's equation is to be used at early times, as the above
analysis of causality shows. The two equations give different results at
times when the front and the wake have not clearly separated. However,
it is useful to be aware of the fact that at such times higher order
effects, or the full Boltzmann equation, could be needed.

We turn next to the second question and explore the relation between
correlation functions and the hydrodynamic equation $Ln=0$, where $L$ is a
differential operator in space-time. Although the computation is classic
textbook material \cite{forester}, the connection with relativistic
causal hydrodynamics has not been noted. The linear response function,
\ie, the expectation value of the retarded density-density correlator,
$\chi$, is given by
\beq
   \chi(k,z) = \chi(0,0) \left[ 1+iz G(k,z) \right],
\label{linear}\eeq
where $z$ is a complex frequency in the upper-half plane, $k$ a spatial
momentum, and $G(k,z)$ is the Green's function, \ie, inverse of the
operator $L$ \cite{forester}. For the diffusion equation one has
\beq
   G(k,z) = \frac{i}{z+iDk^2}\qquad{\rm hence}\qquad
   \chi(k,z) = \frac{iDk^2\chi(0,0)}{z+iDk^2}.
\label{lrf}\eeq
The spectral density is the limit as $k\to0$ of the imaginary part of 
$\chi/k^2$ as $z$ approaches the real line from above. Hence, for the
diffusion equation one finds that
\beq
   \sigma(\omega) = \frac{D\chi(0,0)}{\omega}.
\label{spectral}\eeq
This is the basis for the Green-Kubo formula
\beq
   D\chi(0,0) = \stackrel{lim}{\scriptscriptstyle\omega\to0}
     \omega\sigma(\omega).
\label{kubo}\eeq
Given any microscopic theory (for example, a quantum field theory
specifying the interactions between the tracer particles and the
constituents of the background fluid) which enables one to compute
the spectral function, $\sigma(\omega)$, the diffusion equation can
be obtained if $\omega\sigma(\omega)$ has a finite limit as $\omega$
vanishes. The Green-Kubo relation, along with the positivity of
$\omega\sigma$, provide a derivation of the diffusion equation from the
microscopic theory.  In a complete microscopic computation, the moments
of the spectral density also exist. Since, the spectral density of the
diffusion equation does not have well-defined moments, the diffusion
equation does not care for these details.

A straightforward treatment of Kelly's equation runs into subtleties. From
eq.\ (\ref{reldiffeq}) one sees that
\beq
   G(k,z) = \frac{i}{-i\tau_R z^2+z+iDk^2},\qquad{\rm hence}\qquad
   \sigma(\omega) = \frac{D\chi(0,0)}{\omega(1+\omega^2\tau_R^2)}
    \times \stackrel{lim}{\scriptscriptstyle k\to0}
       \left[1-\frac{\tau_R\omega^2}{k^2 D}\right].
\label{lrfkelly}\eeq
Since the last factor diverges except for constant $M^2=\tau_R\omega^2/Dk^2$,
there seems to be no sensible spectral function otherwise. One could either
proceed with caution along curves of constant $M^2$, or use a regulator.

The text-book regulator is the memory kernel method \cite{forester}. One
delays the response of a system to a disturbance by introducing the memory
kernel---
\beq
   \mathbf J(\x,t) = \int_0^t dt' D(t-t') \nabla n(\x,t'), \qquad
   D(t-t') = - \frac{D}{\tau_R}\exp\left(\frac{t'-t}{\tau_R}\right).
\label{lag}\eeq
where the explicit expression for the memory kernel above constitutes the
relaxation time ansatz.
Putting this together with the continuity equation, one finds
\beq
   G(k,z) = \frac{i}{z+iDk^2/(1-iz\tau_R)},\qquad
   \chi(k,z) = \frac{iD\chi(0,0)k^2}{z+iD(k^2-z^2\tau_R/D)},\qquad
   \sigma(\omega) = \frac{D\chi(0,0)}{\omega(1+\omega^2\tau_R^2)}.
\label{delaysigma}\eeq
Note that this spectral function is perfectly regular as $k\to0$.
Our earlier analysis of dimensionless combinations of variables is also
reflected in this spectral function. To recover the diffusion equation,
the spectral function in (eq.\ \ref{spectral}) must be accurate for
arbitrarily small $\omega$; so this is the limit $K\to0$.  The new term
in the Green's function in eq.\ (\ref{delaysigma}) is a correction to
the diffusion equation to NLO in $K=\omega\tau_R$.  Although the extra
terms are parameterically small in $K$, they have important effects
when $M\ll1$ is constant. These effects arise at $\omega \propto k$,
which is precisely the region at which causal effects manifest themselves.

Since $\omega\sigma(\omega)$ is positive and the Green-Kubo formula
is obeyed for $M\ll1$, this is an acceptable description of diffusion.
The analysis is completed by noting that eq.\ (\ref{lag}) implies that
Fick's law must be replaced by the generalization
\beq
   \left[1+\tau_R\frac{\partial}{\partial t}\right] \mathbf J(\x,t)
         = - D \nabla n(\x,t),
\label{beyondfick}\eeq
which gives the NLO correction in $K$ to the left hand side.  Eliminating
$\mathbf J$ between the continuity equation (eq.\ \ref{continuity}) and
the above relation then yields Kelly's equation (eq.\ \ref{reldiffeq}),
as noticed in \cite{mota}. A phenomenological theory, such as
eq.\ (\ref{beyondfick}), leads to Kelly's equation. However, such a
theory would have two material constants, $D$ and $\tau$,
whereas the microscopic theory has only one.

While the Green-Kubo formula provides the correct reduction of
the microscopic theory to the continuum, the spectral function
in eq.\ (\ref{delaysigma}) contains more. Its first moment
exists, and $\langle\omega\rangle=D/\tau_R$.  Through Fourier
transforms, the moments of the spectral function are related to time
derivatives of the imaginary part of the linear response function,
$\chi''(\x,t;\x',t')=\langle[n(\x,t),n(\x',t')]\rangle/2$, where $n$
is the number operator in the microscopic theory. Each time derivative on
$n(\x,t)$ becomes a commutator with the Hamiltonian. For the first moment,
the double commutator is a generalization of the f-sum rule (also called
the Thomas-Reiche-Kuhn sum rule) of atomic physics, and can be written as
\beq
   \frac1{\chi(0,0)}\int\frac{d\omega}{2\pi}\omega \sigma(\omega) = 
   i \int d\x\left\langle[[n(0,t),H(0,t)],n(\x,t)]\right\rangle.
\label{fsum}\eeq 
A straightforward computation of the right hand side then leads to
$D/\tau_R=\langle v^2\rangle$, the RMS velocity, and reproduces eq.\
(\ref{dexpr}) for the diffusion constant. The f-sum rule knows about the
microscopic constituents through the right hand side of eq.\ (\ref{fsum}),
and hence can be used to constrain models. As an example, note that the
fluctuation-dissipation theorem gives $2mT=D/\tau_R$, where $T$ is the
temperature and $m$ the mass of the diffuser.  For heavy quarks, such
as the charm or bottom, one would expect that a computation of $\tau_R$
would immediately yield $D$ through this relation. For diffusion of
quantum numbers dominated by light quarks, such as the baryon number or
electric charge, where the effective mass of the carriers is unknown,
this relation potentially gives new information.  Higher analogues of the
sum rule exist at higher orders in $K=\omega\tau_R$. Keeping successively
higher moments, one can construct successively higher order terms in the
analogue of Fick's law. This is a construction of effective dynamical
theories at different time scales.  Our second question is now answered.

Finally we consider whether this analysis can be extended to full
hydrodynamics.  The major difference between the diffusion and
Navier-Stokes equations is that the latter are non-linear.  In the
continuum limit, $K\to0$, all hydrodynamic equations can be derived by
combining three kinds of equations: continuity equations (such as eq.\
\ref{continuity}) which relate a density and its current, constitutive
equations (analogues of Fick's law) which relate the current to
the density through a transport coefficient, and thermodynamic
relations of various kinds, which serve to eliminate redundant
variables. Non-linearities arise through the fact that the constitutive
equations contain terms in powers of the velocities.  However, the
dissipative terms are constructed to be linear, and therefore the
essential features discovered here in the diffusion equation generalize
to hydrodynamics.

Corrections to the continuum limit, in the form of NLO terms in $K\ll1$
can then be expected to yield results similar to those we have obtained
for the diffusion equation \cite{mota}. In particular, the memory kernel
method is expected to give correction terms to the constitutive equations
similar to eq.\ (\ref{beyondfick}). As a result, the corrections to the
hydrodynamic limit of the spectral function will be similar to those
encountered in eq.\ (\ref{delaysigma}). All these arguments generalize,
and the NLO corrections to the hydrodynamic equations are always expected
to restore causality. In addition, the Green-Kubo formulae for the
transport coefficients are not modified--- a matter of some importance for
the computation of transport coefficients from QCD using weak-coupling
methods, lattice computations or the AdS/CFT correspondence.  The NLO
corrections in $K$ give rise to a connection between the relaxation time
and transport coefficient through an f-sum rule. Since the f-sum rule
constrains the microscopic degrees of freedom, they can be valuable in
interpreting results obtained from lattice or AdS/CFT techniques. Such a
systematic microscopic derivation of the hydrodynamic equations improves
upon the Israel-Stewart formulation by relating various new material
constants to the old ones through the f-sum rules.

In conclusion, the systematic expansion of transport theory in the
Knudsen number, $K$, which gives the usual hydrodynamic theory to LO in
$K$, \ie, at ${\cal O}(1)$, lacks relativisitic causality at early times
\cite{acausal,kelly}. This acausal region is necessarily shielded at next
to leading order in $K$, \ie, at ${\cal O}(K)$ through the generation
of a time scale, $\tau_R$, and the breakdown of the equations for
$\tau\simeq\tau_R$. Phenomenological theories can give the same causal
hydrodynamic equations as the NLO expansion while introducing seemingly
new material constants.  The systematic expansion in $K$ relates the
new constants to the usual transport coefficients through f-sum rules.
An order-by-order matching of any microscopic computation of a correlation
function to the hydrodynamic equations is possible, and can be seen as
the construction of successive effective long-time theories.

I would like to thank Rajeev Bhalerao and Jean-Yves Ollitrault for
discussions, and IFCPAR project 3104-3 for making the latter possible.


\begin{thebibliography}{99}
\bibitem{eckart}
   C.\ Eckart, {\sl Phys.\ Rev.\/}, 58, 919, (1940).
\bibitem{landau}
   L.\ D.\ Landau and E.\ M.\ Lifshitz, {\sl Fluid Mechanics\/},
    Elsevier, New Delhi (2005).
\bibitem{acausal}
   H.\ D.\ Weymann, {\sl Am.\ J.\ Phys.\/}, 35, 488, (1967);\\
   N.\ G.\ van Kampen, {\sl Physica\/}, 46, 315, (1970).
\bibitem{kelly}
   D.\ C.\ Kelly, {\sl Am.\ J.\ Phys.\/}, 36, 585, (1968).
\bibitem{causal}
   W.\ Israel, \annp, 100, 310, (1976);\\
   J.\ M.\ Stewart, {\sl Proc.\ Roy.\ Soc.\/}, A 357, 59, (1977);\\
   W.\ Israel and J.\ M.\ Stewart, \annp, 118, 341, (1979);\\
   W.\ A.\ Hiscock and L.\ Lindblom, \annp, 151, 466, (1983).
\bibitem{hic}
   A.\ Muronga, \prl, 88, 062302, (2002) and \prc, 69, 034903, (2004);\\
   U.\ Heinz, H.\ Song and A.\ K.\ Chaudhuri, \prc, 73, 034904, (2006);\\
   R.\ Baier, P.\ Romatschke and U.\ A.\ Wiedemann, \prc, 73, 064903 (2006),
   and \npa, 782, 313 (2007);\\
   P.\ Romatschke and U.\ Romatschke, arXiv:0706.1522;\\
   R.\ Bhalerao and S.\ Gupta, arXiv:0706.3428.
\bibitem{grad}
   H.\ Grad, {\sl Phys.\ Fluids\/}, 6, 147, (1963).
\bibitem{forester}
   D.\ Forester, {\sl Hydrodynamics, Correlation Functions and Broken
   Symmetry\/}, W.\ A.\ Benjamin Inc.\/, Reading, MA, USA, 1975.
\bibitem{mota}
   T.\ Koide, G.\ S.\ Denicol, Ph.\ Mota and T.\ Kodama, 
   \prc, 75, 034909, (2007).
\end{thebibliography}
\end{document}